\documentclass[dvips,12pt,a4paper,twoside]{article}
\usepackage{authblk}
\usepackage{amsmath,amsfonts,euscript,amssymb,amsthm}

\usepackage{ifthen}
\usepackage{graphicx}
\usepackage{epsfig}
\usepackage{nicefrac}
\usepackage{mathrsfs}
\usepackage[left=30mm,right=28mm,top=26mm,bottom=25mm]{geometry}
\newcommand{\R}{\mathbb{R}}
\newcommand{\N}{\mathbb{N}}
\newcommand{\C}{\mathbb{C}}

\newcommand{\eps}{\varepsilon}

\def\ri{{\rm i}}%

\def\brem{\begin{remark}}%
\def\erem{\end{remark}}%
\def\beq{\begin{equation}}%
\def\eeq{\end{equation}}%

\def\Vfb{V_{\text{FB}}}
\def\pa{\partial}

\renewcommand{\Re}{\mathrm{Re}}
\renewcommand{\Im}{\mathrm{Im}}

\def\XXint#1#2#3{{\setbox0=\hbox{$#1{#2#3}{\int}$}
     \vcenter{\hbox{$#2#3$}}\kern-.5\wd0}}

\usepackage[
    bookmarks,
    colorlinks=true,
    bookmarksopen=true,
    bookmarksopenlevel=1,
    linkcolor=black,
    citecolor=blue,
    urlcolor=blue,
    filecolor=blue,
    anchorcolor=red,
    pdfstartview=FitH,
    pdfpagelayout=OneColumn,
    plainpages=false, % zur korrekten Erstellung der Bookmarks
    pdfpagelabels, % zur korrekten Erstellung der Bookmarks
    hypertexnames=false, % zur korrekten Erstellung der Bookmarks
    linktocpage, % Seitenzahlen anstatt Text im Inhaltsverzeichnis verlinken
]{hyperref}

\newtheorem{remark}{Remark}[section]

\numberwithin{equation}{section}

\title{Traveling Solitary Waves in the Periodic Nonlinear Schr\"odinger Equation with Finite Band Potentials}
 \author{Tom\'a\v{s} Dohnal \thanks{Technische
     Universit\"at Dortmund, Fakult\"at f\"ur Mathematik,
     Vogelpothsweg 87, D-44227 Dortmund, Germany.}}

\date{August 15, 2013}
\begin{document}

% \end{center}

\maketitle

\begin{abstract}
The paper studies asymptotics of moving gap solitons in nonlinear periodic structures of finite contrast (``deep grating'') within the one dimensional periodic nonlinear Schr\"odinger equation (PNLS). Periodic structures described by a finite band potential feature transversal crossings of band functions in the linear band structure  and a periodic perturbation of the potential yields new small gaps. Novel gap solitons with $O(1)$ velocity despite the deep grating are presented in these gaps. An approximation of gap solitons is given by slowly varying envelopes which satisfy a system of generalized Coupled Mode Equations (gCME) and by Bloch waves at the crossing point. The eigenspace at the crossing point is two dimensional and it is necessary to select Bloch waves belonging to the two band functions. This is achieved by an optimization algorithm. Traveling solitary wave solutions of the gCME then result in nearly solitary wave solutions of PNLS moving at an $O(1)$ velocity across the periodic structure. A number of numerical tests are performed to confirm the asymptotics.
\end{abstract}

%\begin{keywords}

%\begin{AMS}
%\end{AMS}
%35Q55 NLS-like equations (nonlinear Schrödinger) 
%35C20 Asymptotic expansions
%37L60 Lattice dynamics (dyn. syst.)
%33E05 Elliptic functions and integrals 
%41A60 Asymptotic approximations, asymptotic expansions

\pagestyle{myheadings} 
\thispagestyle{plain} 
\markboth{T. DOHNAL}
%{\textsc{Continuation and Stability of Surface Gap Solitons}}
{\textsc{Traveling Solitary Waves in Finite Band Potentials}}

\section{Introduction}\label{S:intro}

Coherent pulses traveling across nonlinear periodic structures like, e.g., optical pulses in  photonic crystals, are of interest from a phenomenological as well as applied point of view. A special case of pulses in periodic structures are gap solitons, which are localized coherent waves with their frequency (or propagation constant) inside a spectral gap of the corresponding linear spatial spectral problem. In certain cases gap solitons exist in families parametrized by velocity so that tuning of the velocity at one fixed frequency becomes possible. 

In the vicinity of spectral edges gap solitons can be approximated by asymptotic slowly varying envelope approximations. The approximation consists of slowly varying envelopes multiplying Bloch waves belonging to the spectral edge. 

In periodic structures with a \textit{finite} (rather than infinitesimal) \textit{contrast} a spectral edge is defined by one or more extrema of the band structure so that the corresponding Bloch waves have zero group velocity. Finite contrast structures are sometimes referred to as ``deep grating'' structures \cite{dSSS96}. The scaling of the envelopes in the generic case of locally parabolic band structure extrema is
%\beq\label{E:GS_scale_parab}
\[
\sqrt{\eps}A(\sqrt{\eps} x, \eps t),
\]
%\eeq
where $1\gg\eps>0$ is the asymptotic parameter, see \cite{BSTU06} and Sec. 2.4.1 of \cite{Pelinov_2011} for 1D examples and \cite{DPS09,DU09} for 2D examples including rigorous justification of the asymptotics. The authors of \cite{BSTU06} consider the 1D nonlinear wave equation with periodic coefficients while in Sec. 2.4.1 of \cite{Pelinov_2011}  and in \cite{DPS09,DU09} the periodic nonlinear Schr\"odinger equation is studied. The scaling of the envelopes and the linear part of the effective equations for the envelopes are, however, determined only by the local shape of the band structure. In both cases the effective equations are either scalar or coupled cubically nonlinear Schr\"odinger equations in the variables $(X,T):=(\sqrt{\eps} x, \eps t)$ with constant coefficients. Localized solitary wave solutions of the effective equations with velocity $v$ in the $(X,T)$-variables produce gap soliton approximations with only the infinitesimal velocity $\sqrt{\eps}v$ in the original $(x,t)$-variables due to the slower scaling of time than space. The approximation is valid on time intervals of length $O(\eps^{-1})$.

In \textit{small} (infinitesimal) \textit{contrast} structures the asymptotic situation for gap solitons is, however, different. Note that periodicity with infinitesimal contrast (here denoted by $\eps$) can open spectral gaps only in 1D problems due to the overlapping of band functions in higher dimensions. In 1D infinitesimal structures gaps open from points where band functions intersect in a transversal manner so that the Fourier waves at $\eps=0$ at the intersection points have nonzero group velocity $\pm c_g$. These Fourier waves then play the role of carrier waves in the asymptotics for gap solitons. The nonzero group velocity also results in equal scaling of time and space in the corresponding envelopes $A_{1,2}$, namely
\[\sqrt{\eps}A_{1,2}(\eps x, \eps t),\]
and the effective model for the envelopes $A_{1,2}(X,T)$, where $(X,T):=(\eps x, \eps t)$,  is a system of two first order equations, so called Coupled Mode Equations (CMEs) 
\beq\label{E:CME_old}
\begin{split}
\ri(\pa_TA_1 + c_g A_1) + \kappa A_2 + \alpha(|A_1|^2+2|A_2|^2)A_1&=0\\ 
\ri(\pa_TA_2 - c_g A_2) + \kappa A_1 +\alpha(|A_2|^2+2|A_1|^2)A_2&=0
\end{split}
\eeq
with $\kappa,\alpha\in \R$. The derivation and justification of \eqref{E:CME_old} for the case of a 1D nonlinear wave equation is in \cite{SU01,GWH01} and for the 1D periodic nonlinear Schr\"odinger equation in \cite{PSn07}. Choosing a moving solitary wave solution \cite{AW89} of \eqref{E:CME_old}, the resulting gap soliton approximations have $O(1)$ velocity due to the equal scaling of time and space in $A_{1,2}(\eps x, \eps t)$. The approximation is, once again, valid on time intervals of length $O(\eps^{-1})$.

Besides the above asymptotic approximations on long but finite time intervals one can also seek exact moving solitary waves. Moving breathers have been considered in \cite{PS08} for the 1D periodic nonlinear Schr\"odinger equation with a small contrast. The resulting breathers travel at an $O(1)$ velocity but could be shown to be localized only on large but finite spatial intervals.

In this paper we consider the asymptoptics on large but finite time intervals and show that an analogous asymptotic situation to that of the infinitesimal contrast can arise also in 1D structures with finite contrast. As we explain, periodic perturbations of so called \textit{finite band potentials} are a suitable example. Such perturbations lead to the opening of gaps from transversal crossings of band functions. In these gaps one can approximate gap solitons via a slowly varying envelope ansatz. The effective  equations are shown to be a generalization of the classical CMEs \eqref{E:CME_old} from the infinitesimal contrast case. Families of localized solutions parametrized by velocity can be constructed by a homotopy continuation from explicitly known solutions of \eqref{E:CME_old}. The resulting gap soliton approximations travel at a wide range of $O(1)$-velocities across a periodic structure of finite contrast. The analysis is performed for the 1D Gross-Pitaevskii equation / periodic nonlinear Schr\"odinger equation (PNLS)
\beq\label{E:PNLS}
\ri \partial_t u + \partial_x^2 u -V(x) u - \sigma |u|^2u=0,
\eeq
where $\sigma \in \R$ and 
\[V(x) = \Vfb(x)+\eps W(x)\]
with a (real) finite band potential $\Vfb$ of period $d>0$ and a piecewise continuous real $d$-periodic perturbation $W$.  Nevertheless, the analysis carries over, for example, to the nonlinear wave equation $\pa_t^2u - \pa_x^2u +V(x)u+\sigma u^3=0$ with $u$ real without significant modifications.

The opening of gaps from transversal crossings of band functions (or touching points when band functions are labeled by magnitude) is a crucial ingredient in our construction of gap solitons with $O(1)$ velocity. An analogous situation of touching bands occurs also in higher dimensions, where isolated touching points are called Dirac points. In the context of the 2D PNLS such points were recently studied, e.g., in \cite{PBFMSC07,ANZ09} for a honeycomb periodic structure. The authors do not, however, study gap solitons near the Dirac points but conical diffraction of initially Gaussian beams. The gap solitons studied in \cite{PBFMSC07} are spatial solitons near a standard extremum of a band function. In fact, it is not clear how our construction of one dimensional moving gap solitons can be generalized to higher dimensions.

A difficulty in setting up the asymptotic ansatz is the selection of Bloch waves with the right group velocity at a crossing point of band functions. This is equivalent to constructing Bloch wave families smooth in $k$ across the crossing points. The eigenspace at the crossing points is two dimensional and we propose a simple optimization algorithm for selecting the right linear combination of a given basis.  

The rest of the paper is structured as follows. In Sec. \ref{S:bloch_review} the linear spectral problem $(-\pa_x^2+V(x))\psi=\lambda\psi$ corresponding to the PNLS is reviewed including the relevant results of Bloch theory. An example of a finite band potential and its band structure is then given. In Sec. \ref{S:k-smooth} the problem of selecting Bloch waves smooth in $k$ is discussed and an optimization algorithm for this selection is introduced. Sec. \ref{S:asymp} provides a formal asymptotic analysis of gap solitons and the derivation of the generalized CMEs as effective equations for the slowly varying envelopes.  In Sec.  \ref{S:ampl_constr} solutions of the generalized CMEs are constructed via homotopy from solutions of CMEs. Finally, in Sec. \ref{S:num} numerical tests verify the expected convergence of the asymptotic approximation error with respect to $\eps$ on time scales of $O(\eps^{-1})$.

\section{Review of Bloch waves and finite band potentials}\label{S:bloch_review}

\subsection{Spectral Problem}\label{S:spec}
The linear spectral problem corresponding to \eqref{E:PNLS} is
\beq\label{E:eig_prob}
L\psi=\lambda \psi,
\eeq
where $L := -\pa_x^2+V(x)$. The spectrum of $L$ is purely continuous (Theorem XIII.90 in \cite{RS4}) and consists of intervals (bands) possibly separated by gaps. In detail
\[\sigma(L)=\bigcup_{n\in \N}[s_{2n-1},s_{2n}],\]
where $s_n\in \R$ and $s_{2n-1}<s_{2n}\leq s_{2n+1}$, $\N:=\{1,2,3,\dots\}$. The spectrum as well as the corresponding solutions $\psi$ can be determined from the Floquet-Bloch problem for the Hill's equation \eqref{E:eig_prob}. Texts on Floquet-Bloch theory for the Hill's equation with periodic coefficients include \cite{MagWin66,Eastham,RS4}. The Floquet-Bloch problem is the eigenvalue problem
\beq
\label{E:Bloch_eq}
\begin{split}
L \psi(x,k) &= \omega(k) \psi(x,k), \qquad x\in [0,d)\\
\psi(d,k) &= e^{\ri k d} \psi(0,k)
\end{split}
\eeq
with $k\in (-\pi/d,\pi/d]$. For each $k\in (-\pi/d,\pi/d]$ problem \eqref{E:Bloch_eq} has a countable set of eigenvalues $(\omega_n(k))_{n\in \N}$. We number the eigenvalues according to size. The family $\omega_n: k \mapsto \omega_n(k)$,$n\in \N$ is usually referred to as the band structure and we call the individual $\omega_n(k)$ the band functions. The quasi-periodic solutions $\psi_n(x,k)$ corresponding to $\omega_n(k)$ are called Bloch waves and have the form
\beq\label{E:Bloch_form}
\psi_n(x,k) = p_n(x,k)e^{\ri kx}, \qquad p_n(x+d,k)=p_n(x,k).
\eeq
We normalize $\|\psi_n(\cdot,k)\|_{L^2(0,2d)}=1$. Note that due to the form \eqref{E:Bloch_form} we have $\|\psi_n(\cdot,k)\|^2_{L^2(0,d)}=1/2$. The spectrum of $L$ can be determined via
\[\sigma(L)=\bigcup_{\substack{n\in \N \\ k\in (-\pi/d,\pi/d]}}\omega_n(k).\]

Complex conjugation of \eqref{E:Bloch_eq} produces the symmetry 
\beq\label{E:om_sym}
\omega_n(-k)=\omega_n(k) 
\eeq
for all $k \in (-\pi/d,\pi/d)$, and for simple eigenvalues $\omega_n(k)$ we also have 
\beq\label{E:conj_sym}
\psi_n(x,-k)=\overline{\psi_n}(x,k).
\eeq
As a second order equation, \eqref{E:eig_prob} has only two linearly independent solutions. The even symmetry of $\omega_n(k)$ thus implies that double eigenvalues $\omega_n(k)$ can occur only at $k=k_0\in \{0,\pi/d\}$. This multiplicity happens if two band functions $\omega_n(k)$ and $\omega_{n+1}(k)$ touch. For a touching point at $k_0=0$ there are two linearly independent $d-$periodic Bloch functions and for $k_0=\pi/d$ there are two linearly independent $2d-$periodic Bloch functions, cf. Theorem 2.1 in \cite{MagWin66}. 

We seek periodic potentials with such touching points $k_0$ that after a (local) relabeling of the band functions to $\omega_+(k)$ and $\omega_-(k)$ two Bloch waves with nonzero group velocities $0<\omega_+'(k_0)=-\omega_-'(k_0)$, cf. Sec. \ref{S:k-smooth}. Although a large class of potentials may produce such a scenario, we restrict our attention to the well known finite band potentials. 

\subsection{Finite Band Potentials}
\begin{figure}[h!]
\begin{center}
 \epsfig{figure = 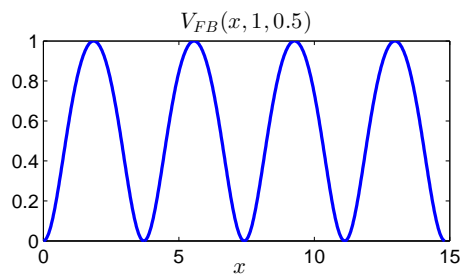,scale=0.60}
 \caption{\label{F:Vfb} \small The finite band potential $\Vfb(x;1,1/2)$.}
\end{center}
\end{figure}

As mentioned above, we wish to study potentials $V$ given by periodic perturbations of finite band potentials. 
A classical example of a finite band potential is
\beq\label{E:jac_ell}
\Vfb(x;m,\eta) = m(m+1)\eta~\text{sn}^2(x;\eta),
\eeq
where $\text{sn}$ is the Jacobi elliptic function, $m\in \R$ and $\eta\in(0,1)$. Equation \eqref{E:eig_prob} with $V=\Vfb$ is usually called Lam\'e's equation. The function $\text{sn}$  is odd and has the period
$2\int_0^1 \frac{dt}{\sqrt{(1-t^2)(1-\eta t^2)}}$.
Therefore, the period of $\Vfb(\cdot;m,\eta)$ is
\[d =\int_0^1 \frac{dt}{\sqrt{(1-t^2)(1-\eta t^2)}}.\]
It is known that Lam\'e's equation has exactly $m$ finite spectral gaps (plus the semi-infinite gap $(-\infty,s_1)$) if and only if $m\in \N$, see \cite{Maier_2008}. In our computations we choose $m=1$ and $\eta= 1/2$. For $\eta = 1/2$ an approximate value of the period is
\[d \approx 3.708149365\]
as provided by \cite{Abra_Steg}. In Fig. \ref{F:Vfb} the function $\Vfb(x;1,1/2)$ is plotted.  
Note that all finite band potentials are analytic, cf. Theorem  XIII.91 (d) in \cite{RS4}.

\begin{figure}[h!]
\begin{center}
 \epsfig{figure = 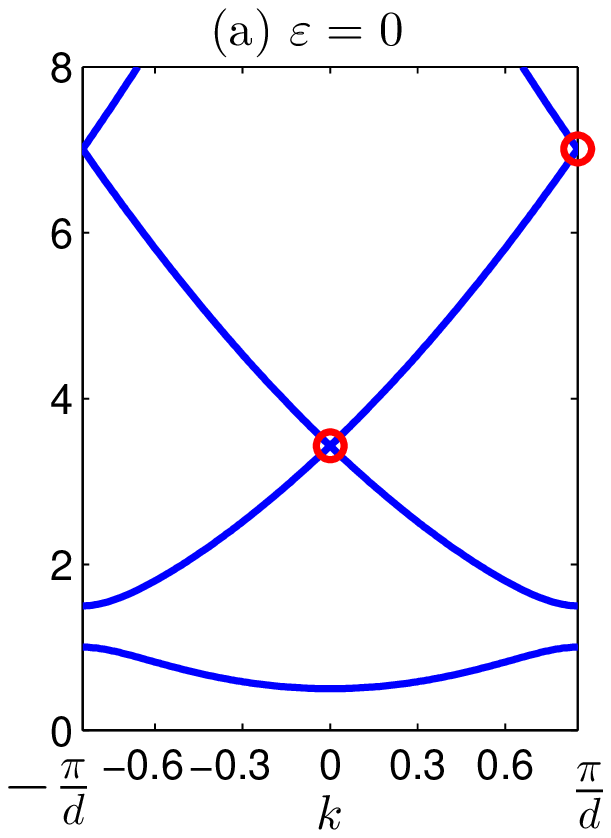,scale=0.60}
\qquad\quad
\epsfig{figure = 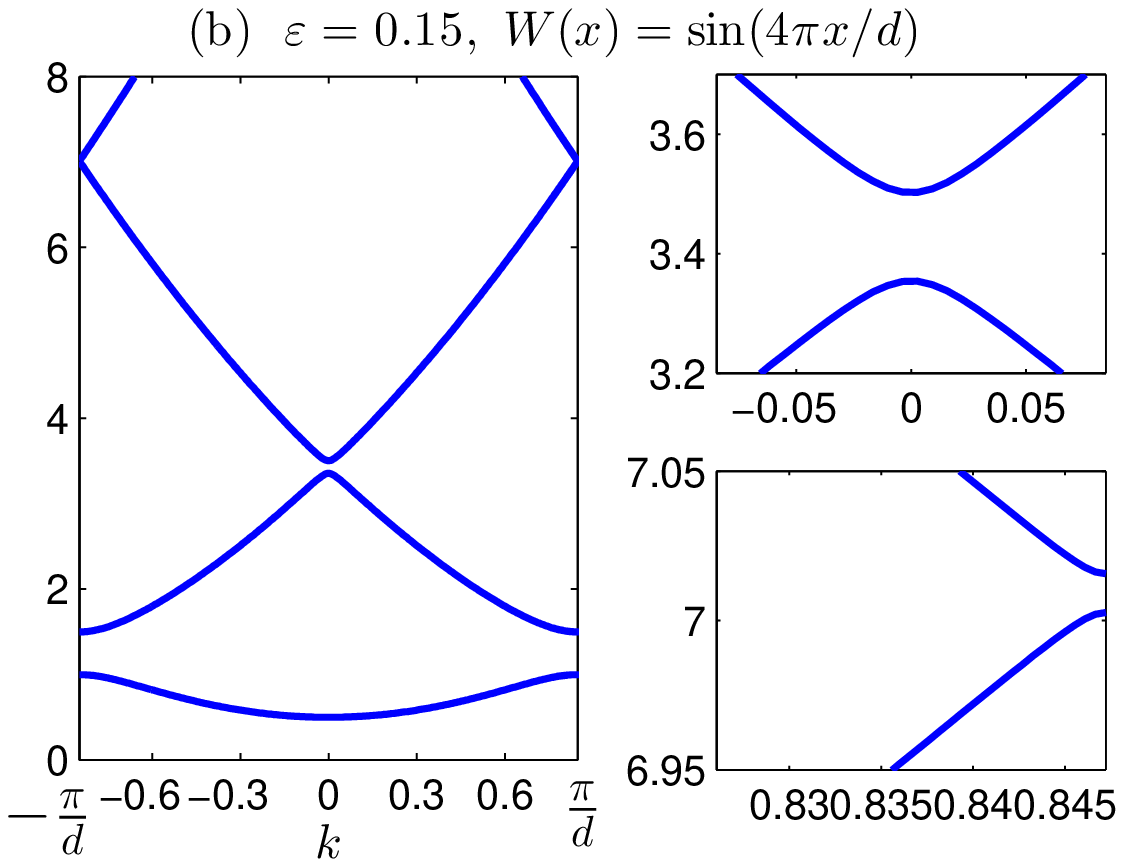,scale=0.60}
 \caption{\label{F:band_str} \small The band structure of \eqref{E:Bloch_eq} for $V(x)=\Vfb(x;1,1/2)+\eps \sin(4\pi x/d)$ with (a) $\eps=0$ and (b) $\eps = 0.15$. In (a) gap bifurcation points are circled. In (b) two regions where gap opening occurs are magnified.}
\end{center}
\end{figure}

The band structure of \eqref{E:Bloch_eq} for $V(x)=\Vfb(x;1,1/2)$ is shown in Fig. \ref{F:band_str} (a), where the sole spectral gap is clearly visible. As the numerical results suggest, under the periodic perturbation $\eps \sin(4\pi x/d)$ new gaps open from the first two touching points of band functions. These points are circled in Fig. \ref{F:band_str} (a) and the new gaps bifurcating from them for $\eps>0$ are magnified in (b). In these computations problem \eqref{E:Bloch_eq} was discretized via the fourth order centered finite difference scheme with $dx=d/1200$.

\section{Smoothness of Bloch Waves in $k$}\label{S:k-smooth}
\begin{figure}[h!]
\begin{center}
 \epsfig{figure = 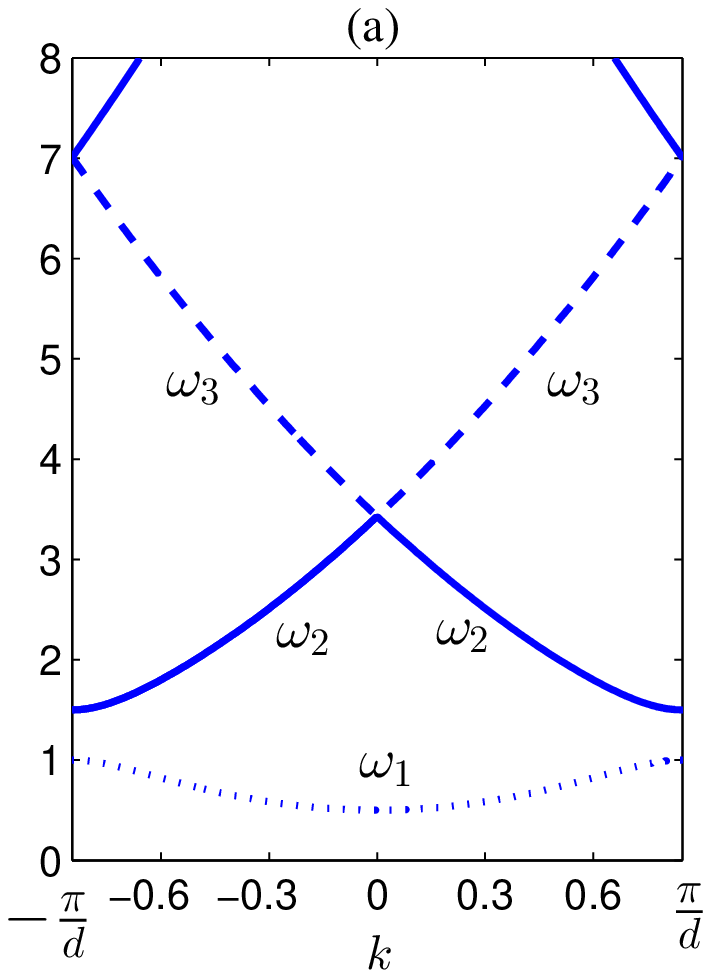,scale=0.60}
\epsfig{figure = 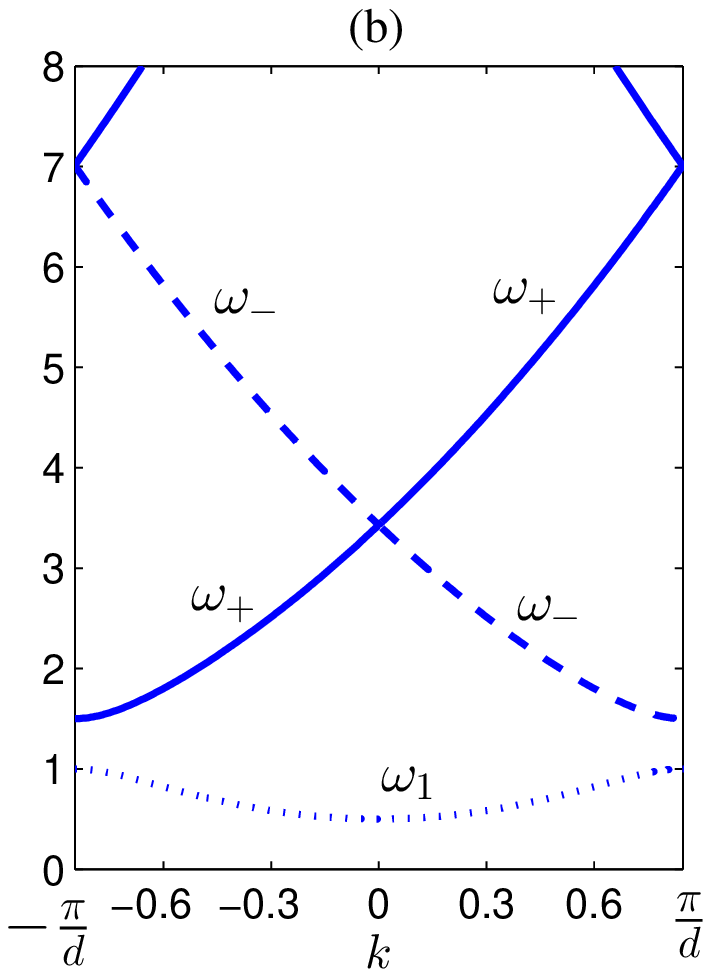,scale=0.60}
\caption{\label{F:band_rename} \small Band structure of \eqref{E:Bloch_eq} for $V(x)=\Vfb(x;1,1/2)$  with eigenvalues labeled according to size in (a) and according to $k-$smoothness in (b).}
\end{center}
\end{figure}

For the asymptotics of gap solitons in gaps bifurcated from touching points of band functions we need to be able to define the group velocity at these points and to find a Bloch wave with this group velocity. The group velocity requires at least differentiability of the band functions. By standard eigenvalue perturbation theory \cite{Kato_1995} eigenvalues $\omega_n(k)$ are analytic in $k$ in regions where they are simple, i.e. away from touching points $k_0$. Clearly, $\omega_n(k)$, which are numbered according to size, are generally not differentiable at touching points, see e.g. $\omega_2(k)$ and $\omega_3(k)$ in Fig. \ref{F:band_rename} (a), which is the example $V=\Vfb(x,1,1/2)$. As shown after Theorem XIII.89 of \cite{RS4}, for the 1D problem \eqref{E:Bloch_eq} with a piecewise continuous $V(x)$ a relabeling is possible so that the resulting band functions are analytic everywhere. For a given touching point $(k,\omega)=(k_0,\omega_0)$ we denote these analytic functions $\omega_+(k)$ and $\omega_-(k)$, where $\pm \omega'_{\pm}(k)\geq 0$. For the touching point $k_0=0,\omega_0\approx3.428$  of the above example the functions $\omega_\pm(k)$ are plotted in Fig. \ref{F:band_rename} (b). For a touching point with $k_0=\pi/d$ the derivatives of $\omega_\pm(k)$ at $k=k_0$ are defined after the $2\pi/d$-periodic extension of $\omega_\pm(k)$.

The Bloch waves $\psi_+(x,k)$ corresponding to $\omega_+(k)$ have nonnegative group velocity and due to symmetry \eqref{E:om_sym} the Bloch waves $\psi_-(x,k)$ corresponding to $\omega_-(k)$ have the opposite group velocity, i.e.
\beq\label{E:cg_sym}
\omega_+'(k)=-\omega_-'(-k) \qquad \text{for all} \ k\in (-\pi/d,\pi/d).
\eeq
Note that the derivatives of $\omega_\pm(k)$ can be expressed as an integral of Bloch waves:
\beq\label{E:cg_id}
\omega_\pm'(k) = -2\ri \int_0^{2d} \pa_x\psi_\pm(x,k)\overline{\psi_\pm}(x,k)dx
\eeq
as follows by the differentiation of $L\psi_\pm(x,k)=\omega_\pm(k) \psi_\pm(x,k)$ in $k$. Symmetry \eqref{E:conj_sym} translates to 
\beq\label{E:conj_sym2}
\psi_+(x,k)=\overline{\psi_-}(x,-k) \qquad \text{for all} \ k\in (-\pi/d,\pi/d)\setminus \{k_0\}.
\eeq

For our gap soliton asymptotics we need to determine the Bloch waves $\psi_\pm(x,k_0)$, which will play the role of carrier waves. In general it is unknown whether Bloch waves can be chosen smooth with respect to $k$ throughout touching points. In Theorem XIII.89 of \cite{RS4} it is shown that for piecewise continuous $V(x)$ continuity with respect to $k$ can be achieved at the touching points. It seems that literature offers no stronger regularity results. For the purposes of our gap soliton asymptotics, we propose the following algorithm to select $\psi_\pm(x,k_0)$.

\subsection{Choice of the Carrier Waves $\psi_\pm(x,k_0)$}\label{S:choice}
At the intersection of the band functions $\omega_\pm(k)$ at $(k,\omega)=(k_0, \omega_0)$ there are two linearly independent solutions $v_1$ and  $v_2$ of \eqref{E:Bloch_eq} with the (quasi-)periodic extension of $v_1$ being even and that of $v_2$ being odd in $x$ \cite{Eastham}. As $k_0\in\{0,\pi/d\}$, the solutions are $2d-$periodic and can be chosen real. We normalize $\|v_{1,2}\|_{L^2(0,2d)}=1$.

The Bloch waves $\phi_1(x):=\psi_+(x,k_0)$ and $\phi_2(x):=\psi_-(x,k_0)$ are specific linear combinations of $v_1$ and $v_2$. We write
\beq\label{E:phi_12}
\phi_1=av_1+bv_2, \qquad \phi_2=cv_1+dv_2.
\eeq
Due to the continuity in $k$ symmetry \eqref{E:conj_sym2} has to hold also at $k=k_0$, i.e. we have $\phi_2=\overline{\phi_1}$ so that
\[c=\bar{a}, \qquad d=\bar{b}\]
and only $a,b\in \C$ need to be determined. We define $(a,b)$ as the solution of
\[\lim_{\delta \to 0} \|\psi_+(\cdot,k_0-\delta)-(av_1+bv_2)\|_{L^2(0,d)}=0.\]
For the purposes of the numerical implementation we fix a small $\delta>0$ (typically $\delta = \tfrac{\pi}{2d}10^{-3}$) and minimize
\[J(a,b):=\|\psi_+^\delta-(av_1+bv_2)\|^2_{L^2(0,d)},\]
where $\psi_+^\delta(x):= \psi_+(x,k_0-\delta)$. This is a four dimensional optimization problem (in the real and imaginary parts of $a$ and $b$) but it can be reduced to an effectively one dimensional problem via the following two constraints. 

Firstly, we require orthogonality of $\phi_1$ and $\phi_2$. 
\[0=\int_0^{2d}\phi_1\overline{\phi_2}dx=\int_0^{2d}\phi_1^2dx=\int_0^{2d}(av_1+bv_2)^2dx = a^2+b^2,\]
where the second equality follows from $\phi_2=\overline{\phi_1}$ and the last equality from the normalization of $v_1$ and $v_2$ and their opposite spatial symmetry such that $v_1v_2$ is an odd function across $x=d$. Consequently result we have
\beq\label{E:ab_sym}
b=s\ri a \qquad \text{with} \quad s\in \{-1,1\},
\eeq
and it remains to determine $a\in \C$ and $s\in \{-1,1\}$.

Secondly, we use the normalization constraint $\|\phi_1\|_{L^2(0,2d)}=1$ so that
\[
\begin{split}
1=&\|\phi_1\|^2_{L^2(0,2d)}=\int_0^{2d}(av_1+s\ri av_2)(\bar{a}v_1-s\ri \bar{a}v_2)dx\\
=& |a|^2 \int_0^{2d} (v_1^2+v_2^2)dx = 2|a|^2
\end{split}
\] 
and
\[a = \frac{1}{\sqrt{2}}e^{\ri \rho}, \qquad \rho \in \R.\]
As a result we need to minimize with respect $\rho \in \R$ and $s\in \{-1,1\}$ the function
\[
\begin{split}
\tilde{J}(\rho,s):=&\|\psi_+^\delta-\frac{1}{\sqrt{2}}e^{\ri\rho}(v_1+s\ri v_2)\|^2_{L^2(0,d)}\\
=& \int_0^d\frac{1}{2}(v_1^2+v_2^2)+|\psi_+^\delta|^2-\sqrt{2}\Re(\psi_+^\delta e^{-\ri\rho}(v_1-s\ri v_2))dx \\
=& 1-\sqrt{2}\Re(\psi_+^\delta e^{-\ri\rho}(v_1-s\ri v_2))dx \\
= & 1-\sqrt{2}\left[\cos(\rho)\int_0^d \Re(\psi_+^\delta (v_1-s\ri v_2))dx+\sin(\rho)\int_0^d \Im(\psi_+^\delta (v_1-s\ri v_2))dx\right],
\end{split}
\] 
where in the one but last equality the normalization of $v_{1,2}$ as well as the property $\|\psi_+^\delta\|^2_{L^2(0,d)}=1/2$, cf. Sec. \ref{S:spec}, have been used.
The derivative of $\tilde{J}$ with respect to $\rho$ vanishes at those values $\rho=\rho_*$, for which 
\[\tan(\rho_*) = \frac{\int_0^d \Im(\psi_+^\delta (v_1-s\ri v_2))dx}{\int_0^d \Re(\psi_+^\delta (v_1-s\ri v_2))dx}.\]
The choice of the optimal $s\in \{-1,1\}$ can be done by simply comparing  $\tilde{J}(\rho_*,s)$ for the two values of $s$. Indeed, the numerics show that one choice of $s$ leads to a near zero value of $\tilde{J}$ while the other choice leads to a value close to $1$.

For purposes of the asymptotics in Sec. \ref{S:asymp} we multiply the families $\psi_+$ and $\psi_-$ (including $\phi_1$ and $\phi_2$) by complex phase factors to ensure 
\beq\label{E:kap_real}
\int_0^{2d}W(x) \phi_1 \overline{\phi_2} dx \in \R,
\eeq
cf. the coefficient $\kappa$ in \eqref{E:CME}.
The phase factors can be determined explicitly. For $\phi_{1,2}$ as defined in \eqref{E:phi_12} we have 
\begin{align*}
\int_0^{2d}W(x) \phi_1 \overline{\phi_2} dx & = \int_0^{2d}W(x) \phi_1^2 dx = \int_0^{2d}W(x) (a^2v_1^2+b^2v_2^2+2abv_1v_2)\\
& = 2ab \int_0^{2d}W(x) v_1v_2 dx = 2s\ri a^2  \int_0^{2d}W(x) v_1v_2 dx,
\end{align*}
due to $b=s\ri a$ and $\int_0^{2d}W(x) v_1^2 dx=\int_0^{2d}W(x) v_2^2 dx$. Defining $\varphi := \arg\left(s\ri a^2\int_0^{2d}W(x) \right.$ $\left. v_1v_2dx\right)$, we, therefore, set
\begin{align*}
\tilde{\psi}_+(x,k) := e^{-\ri \varphi/2}\psi_+(x,k), & \qquad \tilde{\psi}_-(x,k) := e^{\ri \varphi/2}\psi_-(x,k),\\
\tilde{\phi}_1(x) := e^{-\ri \varphi/2}\phi_1(x), & \qquad \tilde{\phi}_2(x) := e^{\ri \varphi/2}\phi_2(x)
\end{align*}
and drop the tildes, so that now \eqref{E:kap_real} holds. Note that the orthogonality, conjugation symmetry and normalization of $\phi_1$ and $\phi_2$ are preserved under the phase shift.

\begin{figure}[h!]
\begin{center}
 \epsfig{figure = 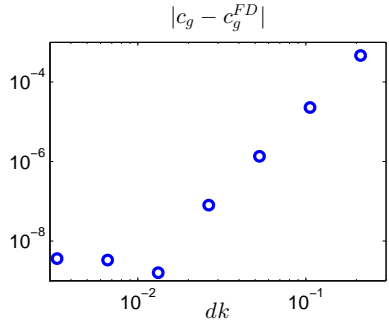,scale=0.60}
\caption{\label{F:cg_conv} \small Convergence of $c_g^\text{FD}$ to $c_g$ in \eqref{E:cg_id2} with respect to the $k$-discretization parameter $dk$.}
\end{center}
\end{figure}

In order to test the algorithm of constructing $\phi_1=\psi_+(x,k_0)$ and $\phi_2=\psi_-(x,k_0)$, we compare the value 
\beq\label{E:cg_id2}
c_g:=\omega_+'(k_0) = -2\ri \int_0^{2d} \pa_x\phi_1(x)\overline{\phi_1}(x)dx
\eeq 
as given by \eqref{E:cg_id} with a finite difference approximation $c_g^\text{FD}$ of $\omega_+'(k_0)$ computed using the fourth order centered finite difference formula with $dk=\frac{\pi}{5*2^nd}$ for $n=0,1,2,\dots,6$. The integral in $c_g$ was approximated using the trapezoidal rule with $dx =d/1200$. Fig. \ref{F:cg_conv} shows a clear convergence of $c_g^\text{FD}$ to $c_g$ as $dk$ decreases.

\section{Asymptotics of gap solitons in narrow gaps}\label{S:asymp}

A generic $d-$periodic perturbation $\eps W$ of a $d-$periodic finite band potential $\Vfb$ in 
\[
V=\Vfb+\eps W 
\]
generates for $\eps>0$ infinitely many gaps in the spectrum $\sigma(-\partial_x^2+V)$, see Theorem 4.6.1 in \cite{Eastham}. Let us assume that one such new gap opens. As explained in Sec. \ref{S:bloch_review}, the opening occurs at the intersection of spectral bands $\omega_+(k)$ and $\omega_-(k)$ at $k=k_0\in \{0,\pi/d\}$. 

Let $\phi_1$ and $\phi_2$ be the Bloch waves constructed in Sec. \ref{S:choice}. They are $2d-$periodic and normalized to satisfy $\|\phi_{1,2}\|_{L^2(0,2d)}=1$.

Analogously to \cite{dSSS96} we propose now the following slowly varying envelope ansatz for gap solitons in the new infinitesimally small spectral gap
\begin{align}\label{E:ans}
u(x,t) \sim & \ \eps^{1/2} \sum_{j=1}^2 A_j(X, T)\phi_j(x)e^{-\ri \omega_0 t} +\eps^{3/2}u_1(x,X,T)e^{-\ri \omega_0 t}\\
& X:= \eps x, \ T :=\eps t \notag
\end{align}
for $\eps\to 0$ with $u_1$ being $2d-$periodic in $x$. Note that the ansatz \eqref{E:ans} is analogous to the case of gap solitons in periodic structures with infinitesimally small contrast, where the carrier Bloch waves $\phi_{1,2}$ are replaced by plane waves \cite{GWH01,SU01}. Substituting \eqref{E:ans} in \eqref{E:PNLS} and collecting equal powers of $\eps$ produces at $O(\sqrt{\eps})$ the linear problems
\[(L-\omega_0)\phi_1=0, \qquad \text{and} \quad (L-\omega_0)\phi_2=0,\]
which are satisfied by the definition of $\phi_{1,2}$. 

At $O(\eps^{3/2})$ we obtain
\begin{align}\label{E:u1_eq}
(L-\omega_0)u_1 = &\left(\ri \phi_1 \pa_T +2\pa_x\phi_1 \pa_X \right)A_1 + \left(\ri \phi_2 \pa_T +2\pa_x\phi_2 \pa_X \right)A_2  \\
 & - W(x) \left(A_1\phi_1+A_2\phi_2\right) - \sigma \left(|A_1|^2A_1|\phi_1|^2\phi_1 +|A_2|^2A_2|\phi_2|^2\phi_2 \right. \notag \\
& \left. + 2|A_1|^2A_2|\phi_1|^2\phi_2+ 2|A_2|^2A_1|\phi_2|^2\phi_1 + A_1^2\overline{A_2}\phi_1^2\overline{\phi_2}+A_2^2\overline{A_1}\phi_2^2\overline{\phi_1}\right). \notag
\end{align}
By Fredholm alternative a $2d-$periodic solution (with respect to $x$) of \eqref{E:u1_eq} exists only if the right hand side is $L^2(0,2d)-$orthogonal to $\phi_1$ and $\phi_2$. 
The resulting two solvability conditions are the \textit{generalized Coupled Mode Equations} (gCMEs)
\beq\label{E:CME}
 \begin{split}
  \ri \left(\pa_T +c_g \pa_X\right)A_1 + \kappa A_2 + \kappa_s A_1 +\alpha (|A_1|^2+2|A_2|^2)A_1 & \\
	   + \beta (2|A_1|^2+|A_2|^2)A_2 + \overline{\beta}A_1^2\overline{A_2} +\gamma A_2^2 \overline{A_1} &= 0,\\
  \ri \left(\pa_T -c_g \pa_X\right)A_2 + \kappa A_1 + \kappa_s A_2 +\alpha (|A_2|^2+2|A_1|^2)A_2 & \\
	    + \overline{\beta} (2|A_2|^2+|A_1|^2)A_1 + \beta A_2^2\overline{A_1} +\overline{\gamma} A_1^2 \overline{A_2} &= 0,\\
 \end{split}
\eeq
where
\begin{align*}
c_g & = \omega_+'(k_0)= -2\ri \int_0^{2d} \phi_1'(x)\overline{\phi_1}(x)dx = 2\ri \int_0^{2d} \phi_2'(x)\overline{\phi_2}(x)dx \in \R,\\
\kappa & = - \int_0^{2d}W(x) \phi_2 \overline{\phi_1} dx  = - \int_0^{2d}W(x) \phi_1 \overline{\phi_2} dx \in \R, \\
\kappa_s & = - \int_0^{2d}W(x) |\phi_1|^2 dx  = - \int_0^{2d}W(x) |\phi_2|^2  dx \in \R, \\
\alpha & = -\sigma\int_0^{2d}|\phi_1|^4dx = -\sigma\int_0^{2d}|\phi_2|^4dx = -\sigma\int_0^{2d}|\phi_1|^2|\phi_2|^2dx \in \R,\\
\beta & = -\sigma\int_0^{2d}|\phi_1|^2 \phi_2\overline{\phi_1}dx = -\sigma\int_0^{2d}|\phi_2|^2 \phi_2\overline{\phi_1}dx \in \C, \\
\gamma & = -\sigma\int_0^{2d} \phi_2^2 \overline{\phi_1}^2 dx \in \C.
\end{align*}
The identities in $c_g, \alpha, \kappa_s$ and $\beta$ follow from $\phi_2=\overline{\phi_1}$. The fact that $\kappa\in \R$ follows from \eqref{E:kap_real} and $c_g\in \R$ can be seen via integration by parts. In fact, our choice $c_g=\omega_+'(k_0)$ ensures $c_g>0.$

\begin{remark}
A system of the same type as \eqref{E:CME} appears in \cite{dSSS96} but no explanation of which periodic structures lead to this effective model is given.
\end{remark}
\begin{remark}
Note that for odd functions $W$, i.e. $W(x)=-W(-x)$ for all $x\in \R$, we get $\kappa_s=0$ due to the following calculation.
\[\int_0^{2d}W(x) |\phi_1|^2 dx =\int_0^{2d}W(x)(|a|^2v_1^2+|b|^2v_2^2)dx + \int_0^{2d}W(x)(a\overline{b}+b\overline{a}) v_1v_2dx\]
The first integral on the right hand side is zero due to the evenness of $v_1^2$ and $v_2^2$ about $x=d$ and the second integral vanishes due to \eqref{E:ab_sym}. 

In fact, one can assume without loss of generality that $\kappa_s=0$ because the terms $\kappa_s A_\pm$ can be removed by the simple gauge transformation $A_\pm(X,T) \rightarrow A_\pm(X,T)e^{\ri \kappa_s T}$.
\end{remark}
\begin{remark}
The above formal asymptotics predict that the correction term $u_1$ is a sum of terms of the form $B(\eps x,\eps t)f(x)$, where $B$ is any of the $(X,T)$-dependent coefficients on the right hand side of \eqref{E:u1_eq}, i.e. $\pa_T A_1, \pa_X A_1, A_2,A_1, |A_1|^2A_1$ etc. Therefore, assuming boundedness of $A_{1,2}$ and its first derivatives, we formally get
\beq\label{E:conv_predict}
\|\eps^{3/2} u_1(\cdot,\eps\cdot,T)\|_{L^2(\R)} = O(\eps).
\eeq
\end{remark}

\section{Construction of Solutions to the Generalized Coupled Mode Equations}\label{S:ampl_constr}

Explicit solutions of \eqref{E:CME} are not known. System \eqref{E:CME} is, however, a generalization of the classical CMEs \eqref{E:CME_old} for envelopes of pulses in the nonlinear wave equation with a periodic structure of infinitesimal contrast \cite{AW89,GWH01}. gCMEs \eqref{E:CME} reduce to \eqref{E:CME_old} if we set $\kappa_s=\beta=\gamma=0$. System \eqref{E:CME_old} has the following explicitly known family of solitary waves, see \cite{AW89,GWH01,DH07},
\begin{equation}\label{E:1D_gap_sol}
\begin{split}
A_1^{(0)} &= \nu a e^{\ri\eta}\sqrt{\frac{|\kappa|}{2|\alpha|}}\sin(\delta)\Delta^{-
1}e^{\ri\nu\zeta} \text{sech}(\theta - \ri\nu\delta/2),\\
A_2^{(0)} &= -a e^{\ri\eta}\sqrt{\frac{|\kappa|}{2|\alpha|}}\sin(\delta)\Delta e^{\ri\nu\zeta} \text{sech}(\theta  +\ri\nu\delta/2),
\end{split}
\end{equation}
where
\begin{align*}
\nu &=\text{sign}(\kappa \alpha), \quad  a = \sqrt{\frac{2(1-v^2)}{3-v^2}}, \quad \Delta = \left(\frac{1-v}{1+v}\right)^{1/4}, \quad e^{\ri\eta} = \left(-\frac{e^{2\theta}+e^{-\ri\nu\delta}}{e^{2\theta}+e^{\ri\nu\delta}}
\right)^{\frac{2v}{3-v^2}}\\
\theta & = \mu \kappa
\sin(\delta)\left(\frac{X}{c_g}-vT\right), \quad \zeta = \mu
\kappa \cos(\delta)\left(\frac{v}{c_g}X-T\right), \quad \mu = (1-v^2)^{-1/2}
\end{align*}
with the parameters of velocity $v\in (-1,1)$ and ``detuning'' $\delta \in [0,\pi]$. 

For $\delta =\pi/2$ we get $\zeta=0$ and both $A_1^{(0)}$ and $A_2^{(0)}$ depend only on the moving frame variable $\tilde{\theta}=\theta(\delta=\pi/2) = \mu \kappa \left(\frac{X}{c_g}-vT\right)$. For this case we carry out a numerical homotopy continuation of solutions \eqref{E:1D_gap_sol} to solutions of \eqref{E:CME} with $\kappa_s,\beta $ and $\gamma$ nonzero. In the moving frame variable $\tilde{\theta}$ system \eqref{E:CME}  reads 
\beq\label{E:CME_mov_frame}
 \begin{split}
  \ri a_+ \pa_{\tilde{\theta}} A_1 + \kappa A_2 + \kappa_s A_1 +\alpha (|A_1|^2+2|A_2|^2)A_1 & \\
	   + \beta (2|A_1|^2+|A_2|^2)A_2 + \overline{\beta}A_1^2\overline{A_2} +\gamma A_2^2 \overline{A_1} &= 0,\\
  \ri a_- \pa_{\tilde{\theta}} A_2 + \kappa A_1 + \kappa_s A_2 +\alpha (|A_2|^2+2|A_1|^2)A_2 & \\
	    + \overline{\beta} (2|A_2|^2+|A_1|^2)A_1 + \beta A_2^2\overline{A_1} +\overline{\gamma} A_1^2 \overline{A_2} &= 0,\\
 \end{split}
\eeq
where $a_+=\mu \kappa (1-v)$ and $a_-=-\mu \kappa (1+v)$.

We solve \eqref{E:CME_mov_frame} numerically in the Fourier space via the Petviashvili iteration \cite{AM05} starting at $\kappa_s=\beta = \gamma=0$ with the solution given by \eqref{E:1D_gap_sol} and continuing up to the desired values of $\kappa_s,\beta$ and $\gamma$. We use a single continuation parameter. The computational domain the the $\tilde{\theta}$-variable is chosen large enough to support the localized profiles of $A_{1,2}$.

The coefficients in \eqref{E:CME} are computed by numerically evaluating the corresponding integrals of Bloch functions using the trapezoidal rule with $dx =d/1200$. The Bloch functions are computed by the fourth order centered finite difference scheme with the same value of $dx$.

\begin{figure}[h!]
\begin{center}
 \epsfig{figure = 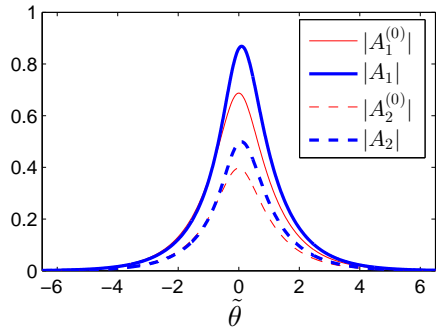,scale=0.6}
\caption{\label{F:CME_prof} \small The solutions $|A_{1,2}|$ of \eqref{E:CME} with $c_g=1, \kappa=0.5, \kappa_s=0,$ $\alpha=1, \beta = 0.3(1+\ri)$ and $\gamma=0.2(1+\ri)$, and the solutions $|A_{1,2}^{(0)}|$  in \eqref{E:1D_gap_sol} with $c_g=1, \kappa=0.5, \alpha=1$  and $\delta=\pi/2, v=0.5$, from which $A_{1,2}$ were computed by homotopy continuation.}
\end{center}
%CME_profile_example.mat
\end{figure}

In Fig. \ref{F:CME_prof} an example of a solution $A_{1,2}$ of \eqref{E:CME} is plotted together with the solution $A_{1,2}^{(0)}$  in \eqref{E:1D_gap_sol}, from which the homotopy continuation was started. Note that while  $|A_{1,2}^{(0)}|$ are even in $\tilde{\theta}$, the profiles $|A_{1,2}|$ are not symmetric. This lack of symmetry is caused by $\Im(\beta)\neq 0$ and $\Im(\gamma)\neq 0$. As easily checked, when  $\Im(\beta)= 0$ and $\Im(\gamma)= 0$, then \eqref{E:CME} has the symmetry $A_{1,2}(\tilde{\theta}) \rightarrow \overline{A_{1,2}}(-\tilde{\theta})$. While if $\Im(\beta)\neq 0$ or $\Im(\gamma)\neq 0$, this symmetry is lost.

\section{Numerical Results on Moving Gap Solitons}\label{S:num}

We compare the asymptotic approximation
\[\sqrt{\eps} u_0(x,t):=\sqrt{\eps}\sum_{j=1}^2 A_j(\eps x, \eps t)\phi_j(x)e^{-\ri \omega_0 t}\]
and a numerical approximation of \eqref{E:PNLS} with the initial condition $u(x,0)=$$\sqrt{\eps}u_0(x,0)$. The integration of  \eqref{E:PNLS} is performed by the second order split step method (Strang splitting), where the PNLS \eqref{E:PNLS} is split according to $\pa_tu = \mathcal{L}u +\mathcal{N}(u)$ with $\mathcal{L}=\ri\pa_x^2$ and $\mathcal{N}(u)=-\ri V(x) u-\ri\sigma|u|^2$. The linear problem $\pa_tu = \mathcal{L}u$ is then solved in Fourier space via fft and the nonlinear problem $\pa_tu = \mathcal{N}(u)$ is solved exactly \cite{WH86}. We use a large computational domain such that the solution tails at the boundary stay below $10^{-5}$ in (relative) amplitude. For the spatial discretization we use $dx=0.1$ and for time stepping $dt=0.01$.   

In both examples below we study gap solitons in the gap bifurcating from the intersection of band functions $\omega_\pm(k)$ at $(k_0,\omega_0) \approx  (0,3.428)$ for potentials $V(x)=\Vfb(x;1,1/2)+\eps W(x)$, see Fig. \ref{F:band_str}. 

\subsection{Odd Perturbation $W(x)$}\label{S:W_odd}
In this example we choose the odd perturbation
\[W(x) = \sin(4\pi x/d)\]
so that $\kappa_s=0$ in \eqref{E:CME}.

Numerically obtained values of the CME coefficients are
\beq\label{E:coeffs}
\begin{split}
&c_g \approx 3.3666546, \ \kappa \approx  0.4934402, \\ 
&\alpha \approx  0.1357055, \ \beta \approx \ri~6.4865707*10^{-7}, \ \gamma \approx  7.2199824*10^{-6}.
\end{split}
\eeq

\begin{figure}[h!]
\begin{center}
 \epsfig{figure = 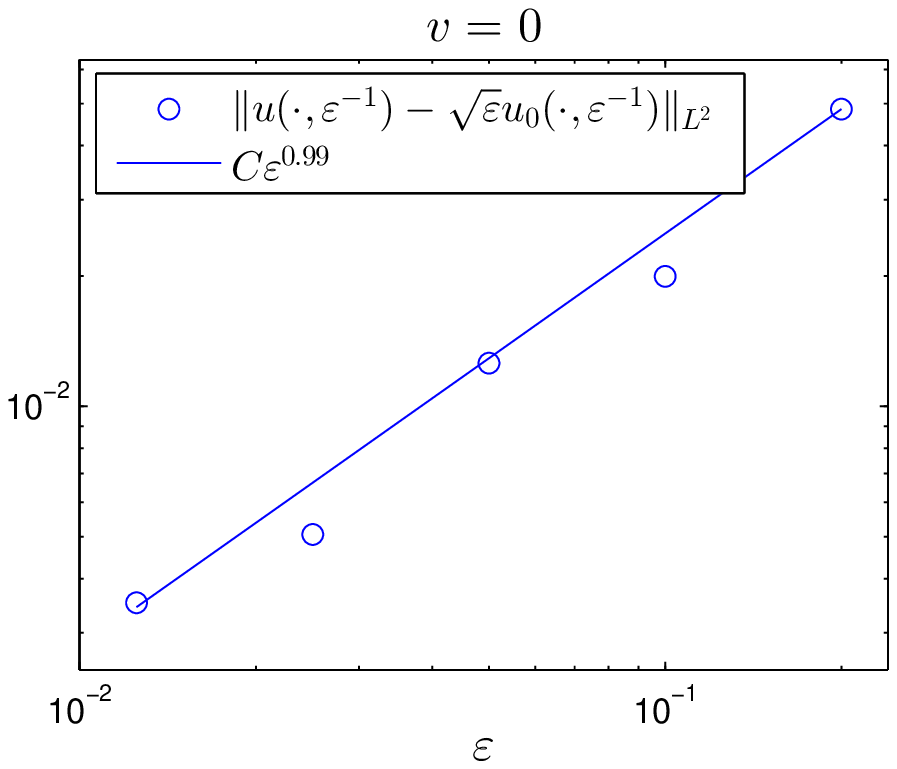,scale=0.60}
\epsfig{figure = 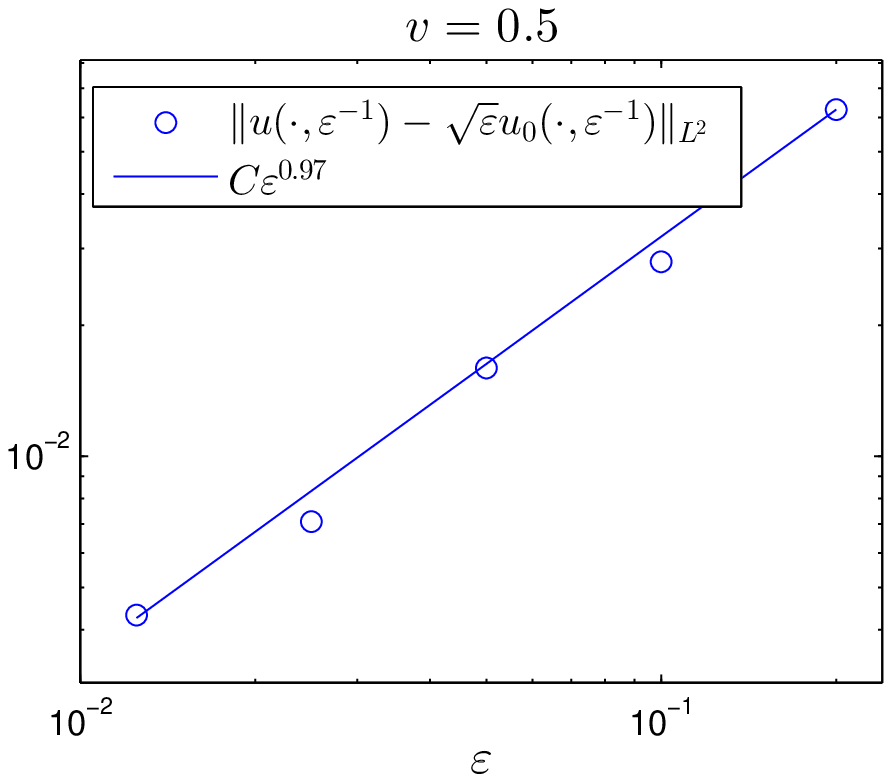,scale=0.60}

\epsfig{figure = 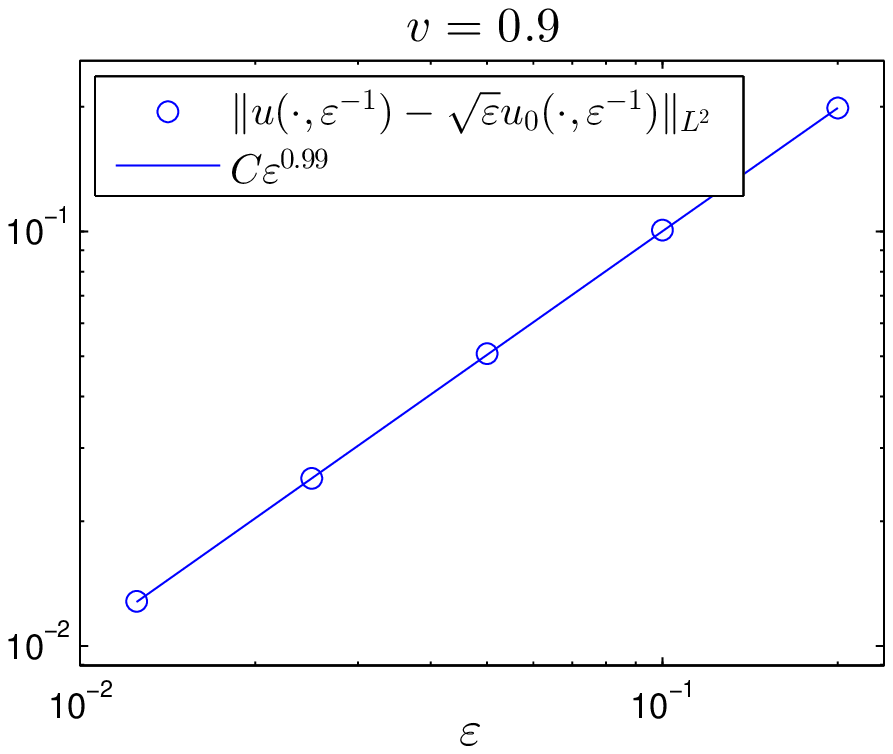,scale=0.60}
\epsfig{figure = 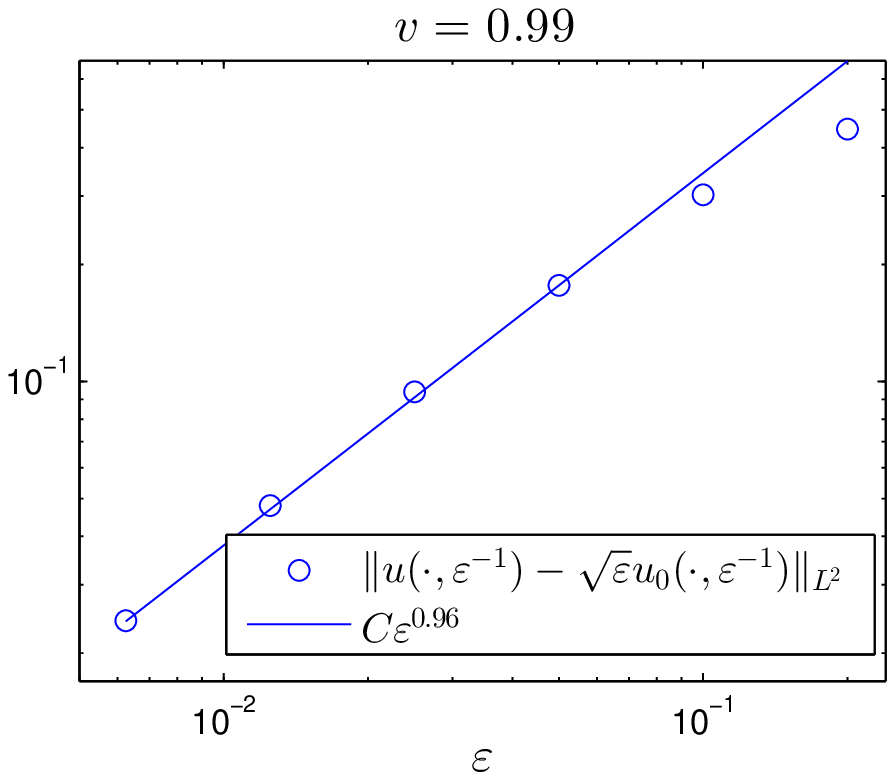,scale=0.60}
 \caption{\label{F:eps_conv} \small Convergence of the $L^2$-error  at $t=\eps^{-1}$, i.e. of $\|u(\cdot,\eps^{-1})-\sqrt{\eps} u_0(\cdot, \eps^{-1})\|_{L^2}$, for the choice of $\Vfb(x;1,1/2)$ given in \eqref{E:jac_ell}  and $W(x)=\sin\left(\tfrac{4\pi x}{d}\right)$, $k_0=0, \omega_0\approx 3.428$. Four velocity parameters in $A_{1,2}$ are tested: $v=0, 0.5, 0.9,$ and $0.99$. The slopes of the full lines are obtained by interpolations of all the computed error values except for $v=0.99$, where only the four smallest values are interpolated.}
\end{center}
%sol_data_PNLS_vel_0_eps_p1_num_envel.mat etc.
%sol_data_PNLS_vel_p5_eps_p1_num_envel.mat  etc.
%sol_data_PNLS_vel_p5_eps_p9_num_envel.mat  etc.
%sol_data_PNLS_vel_p5_eps_p99_num_envel.mat  etc.
\end{figure}

In Fig. \ref{F:eps_conv} the convergence of the $L^2$-error $\|u(\cdot,t)-\sqrt{\eps} u_0(\cdot, t)\|_{L^2}$ at $t=\eps^{-1}$ is shown. The four values of the velocity parameter $v\in\{0,0.5,0.9,0.99\}$ in the moving frame variable $\tilde{\theta}$ are tested. In all cases the observed convergence rate is close to $\eps^1$ as predicted by the formal asymptotics \eqref{E:conv_predict}. Note that the relative error is of the same asymptotic order as the absolute error due to the fact that the error consists of terms of the form $\sqrt{\eps}f(\eps x)g(x)$ and because $\|\sqrt{\eps}f(\eps \cdot)\|_{L^2(\R)} = \|f\|_{L^2(\R)}$ for all $\eps>0$.

\begin{figure}[h!]
\begin{center}
 \epsfig{figure = 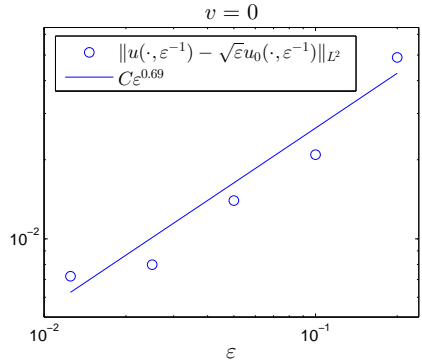,scale=0.6}
\caption{\label{F:eps_conv_ignore_beta} \small The error convergence as in Fig. \ref{F:eps_conv} for the case $v=0$ but ignoring $\beta$ and $\gamma$ (setting $\beta=\gamma=0$) in the gCMEs \eqref{E:CME}. Clearly, a slower convergence and larger error values are produced.}
\end{center}
%sol_data_PNLS_vel_0_eps_p1.mat.mat etc.
\end{figure}

Although the values of $\beta$ and $\gamma$ in \eqref{E:coeffs} are very small, a numerical test confirms that they are nonzero. Indeed, setting $\beta=\gamma=0$ in the computation of the envelopes $A_{1,2}$ and studying once again $\|u(\cdot,\eps^{-1})-\sqrt{\eps} u_0(\cdot, \eps^{-1})\|_{L^2}$ with the new initial data $u(x,0)=\sqrt{\eps}u_0(x,0)$ produces a slower convergence rate, approximately $\eps^{0.69}$, see Fig. \ref{F:eps_conv_ignore_beta}.

\begin{figure}[h!]
\begin{center}
 \epsfig{figure = 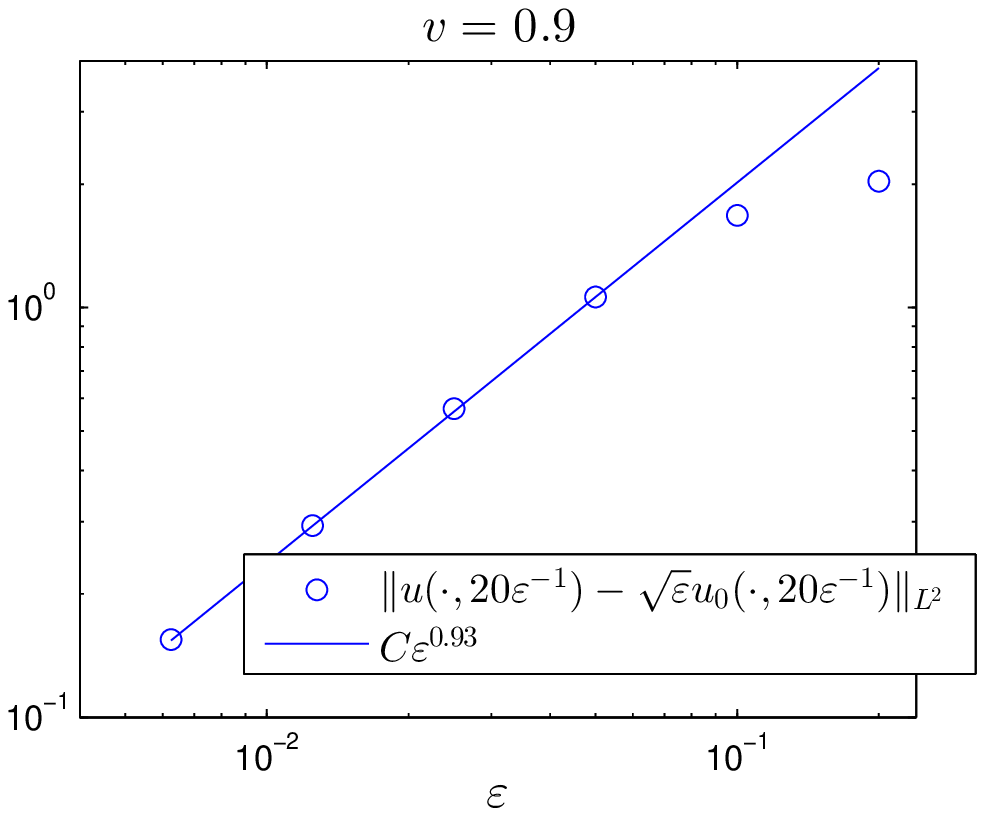,scale=0.6}
\caption{\label{F:eps_conv_LG_time} \small The error convergence for the same setting as in Fig. \ref{F:eps_conv} and $v=0.9$ at the large time $t=20\eps^{-1}$.}
\end{center}
%sol_data_PNLS_vel_p9_eps_p1_num_envel_LG.mat   etc.
\end{figure}

In order to test the behavior at very large times, the numerical error analysis has been performed for $v=0.9$ also at $t=20\eps^{-1}$ with the resulting convergence rate $O(\eps^{0.93})$, see Fig. \ref{F:eps_conv_LG_time}. 

\begin{figure}[h!]
\begin{center}
\epsfig{figure = 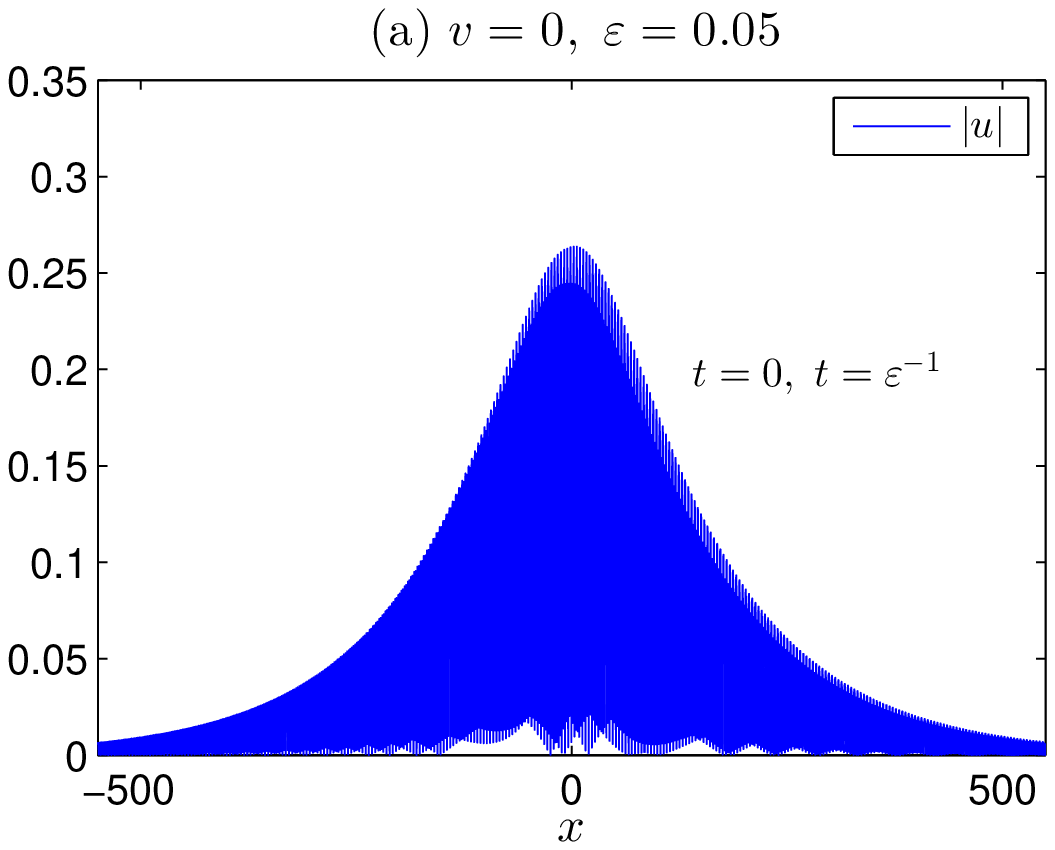,scale=0.5}
\epsfig{figure = 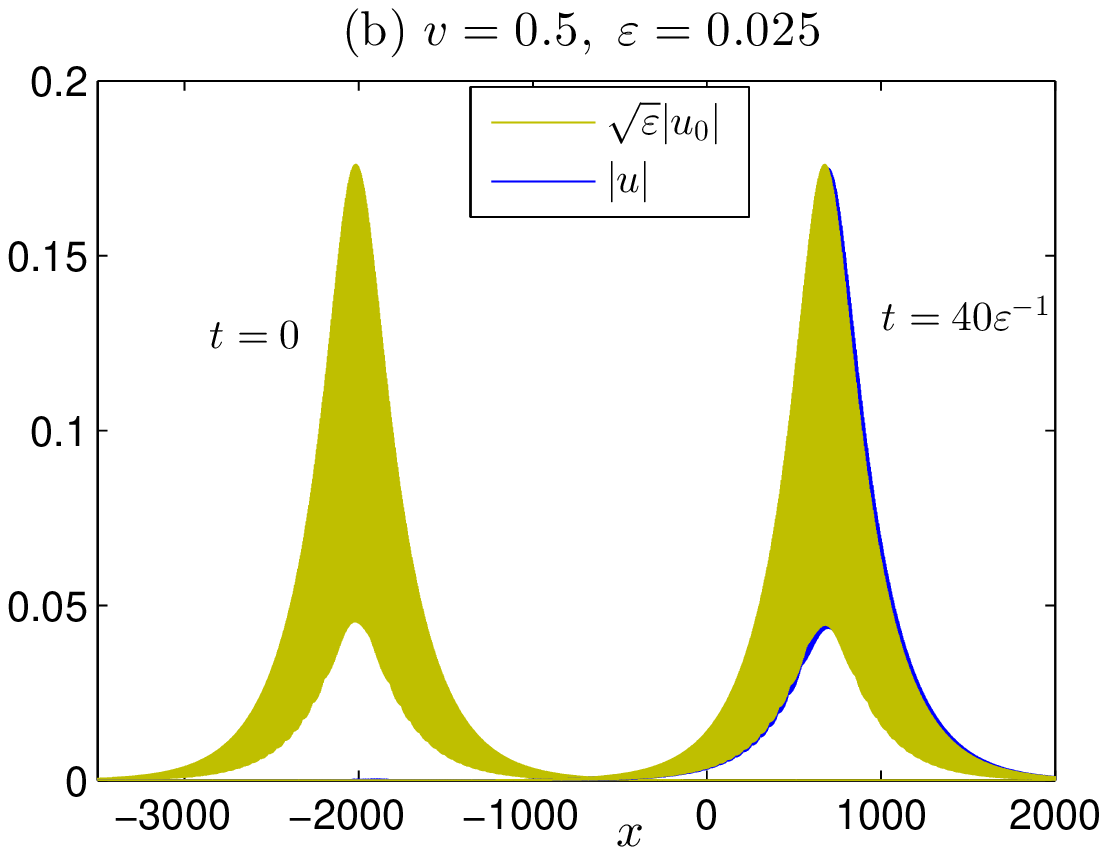,scale=0.5}
\epsfig{figure = 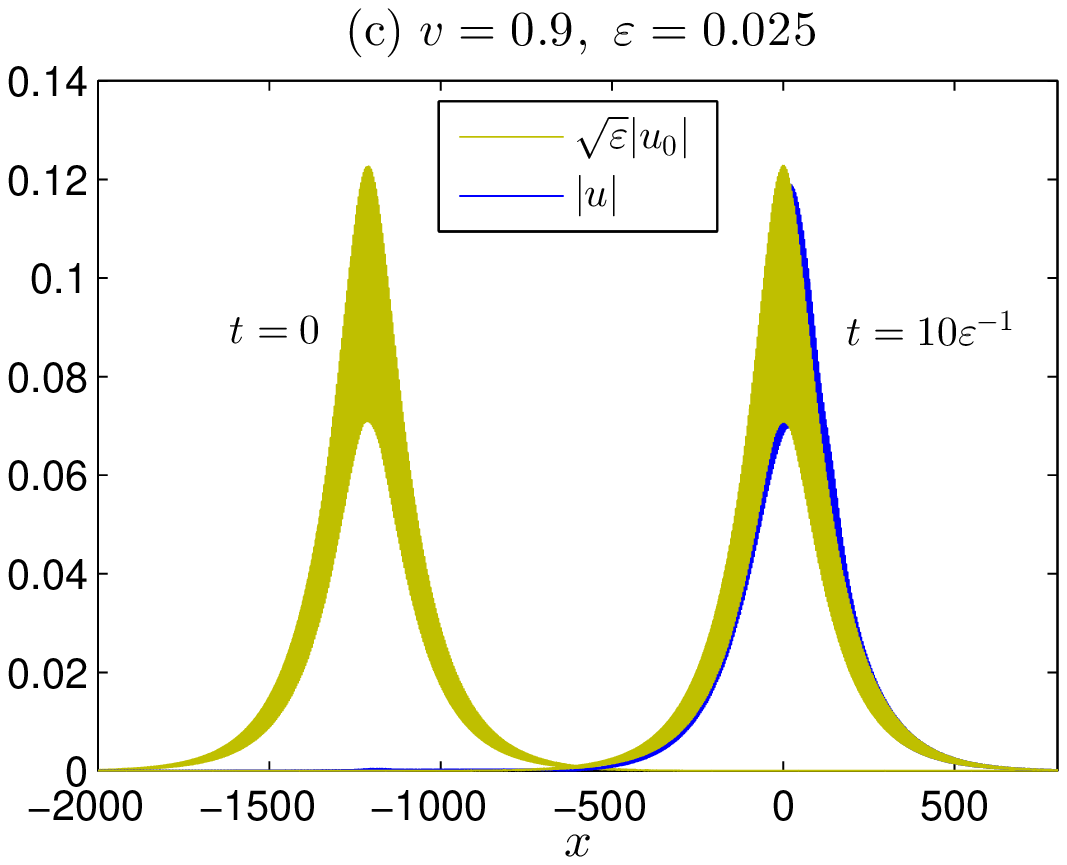,scale=0.5}
\epsfig{figure = 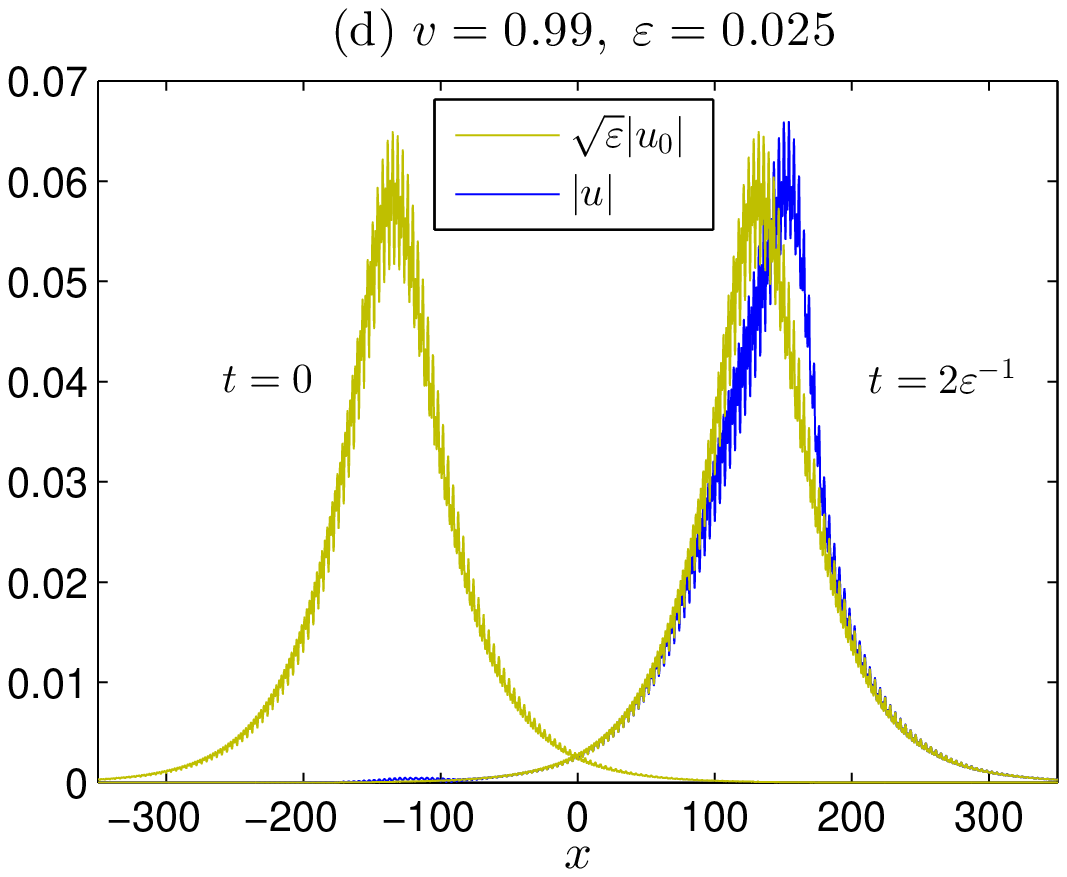,scale=0.5}
\caption{\label{F:sol_plots} \small The numerical solution $|u|$ and the asymptotic solution $\sqrt{\eps}|u_0|$ at $t=0$ and at a time $t=O(\eps^{-1})$ for the same setting as in Fig. \ref{F:eps_conv}.  In (a) the approximation $\sqrt{\eps}|u_0|$ at $t=0$ as well as at $t=\eps^{-1}$ are optically indistinguishable from $|u(x,\eps^{-1})|.$}
\end{center}
%sol_data_PNLS_vel_0_eps_p025_num_envel.mat  
%sol_data_PNLS_vel_p99_eps_p025_num_envel.mat  
\end{figure}

Fig. \ref{F:sol_plots} shows the modulus of the numerical solution $u$ as well as of the asymptotic approximation $\sqrt{\eps}|u_0|$ at $t=0$ and a time $t=O(\eps^{-1})$ for the four velocity parameters $v$ from Fig \ref{F:eps_conv}. For negative values of $v$ the modulus profiles are identical. Note that mainly for small values of $v$ the large width of the pulses compared to the period of $V$ makes it difficult to visually distinguish individual oscillations in $u$ and $u_0$. 

\subsection{Even Perturbation $W(x)$}
\begin{figure}[h!]
\begin{center}
\epsfig{figure = 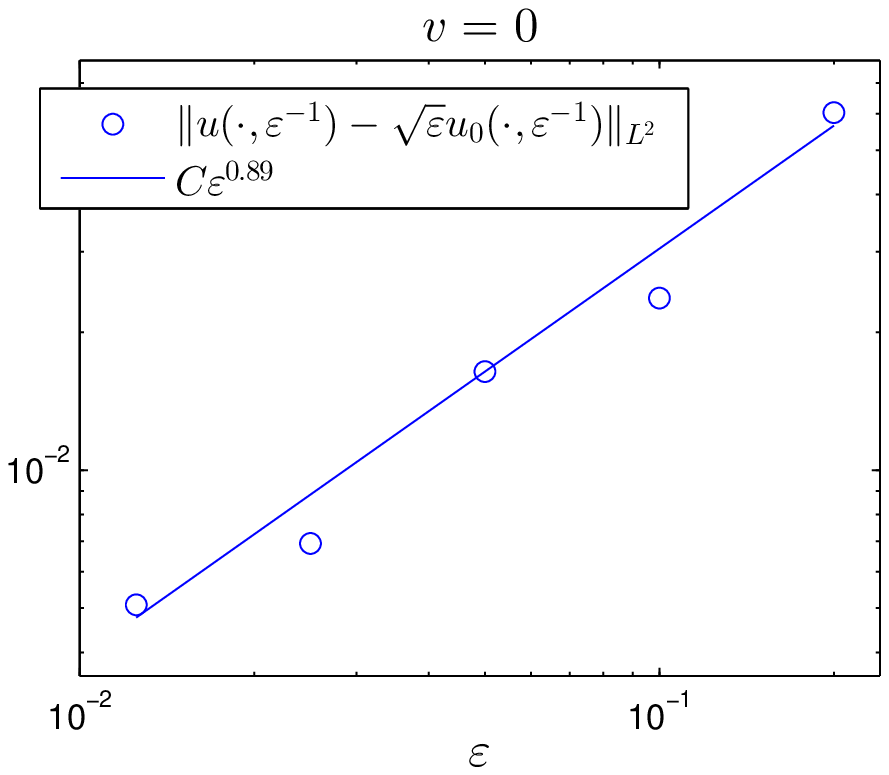,scale=0.60}
\epsfig{figure = 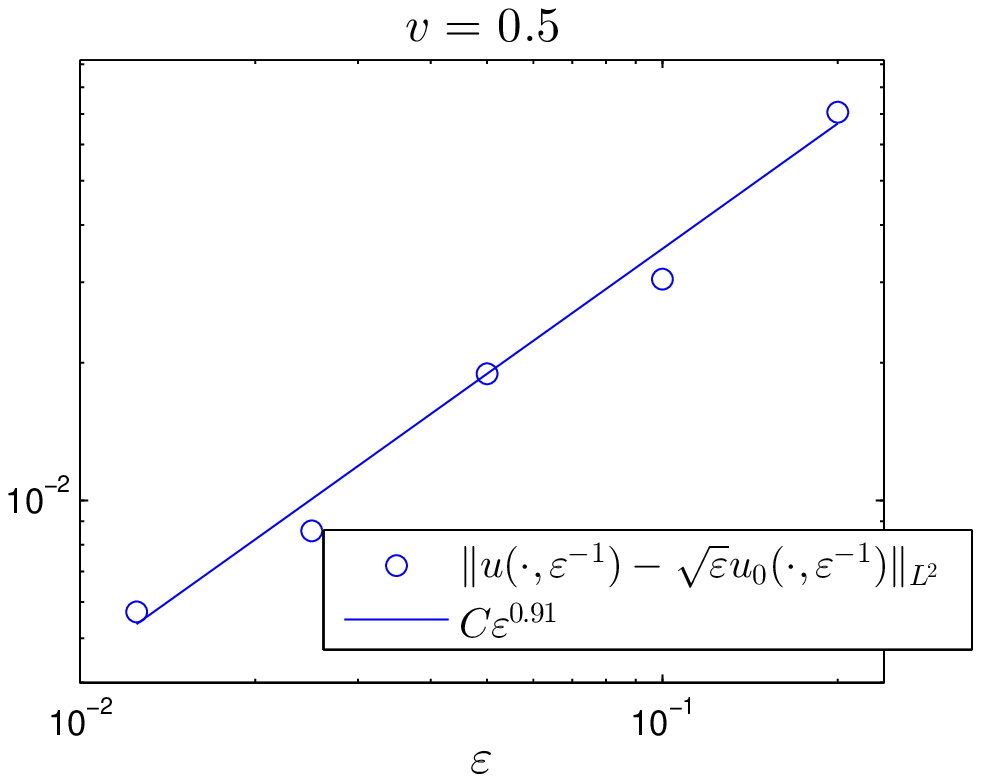,scale=0.60}
 \epsfig{figure = 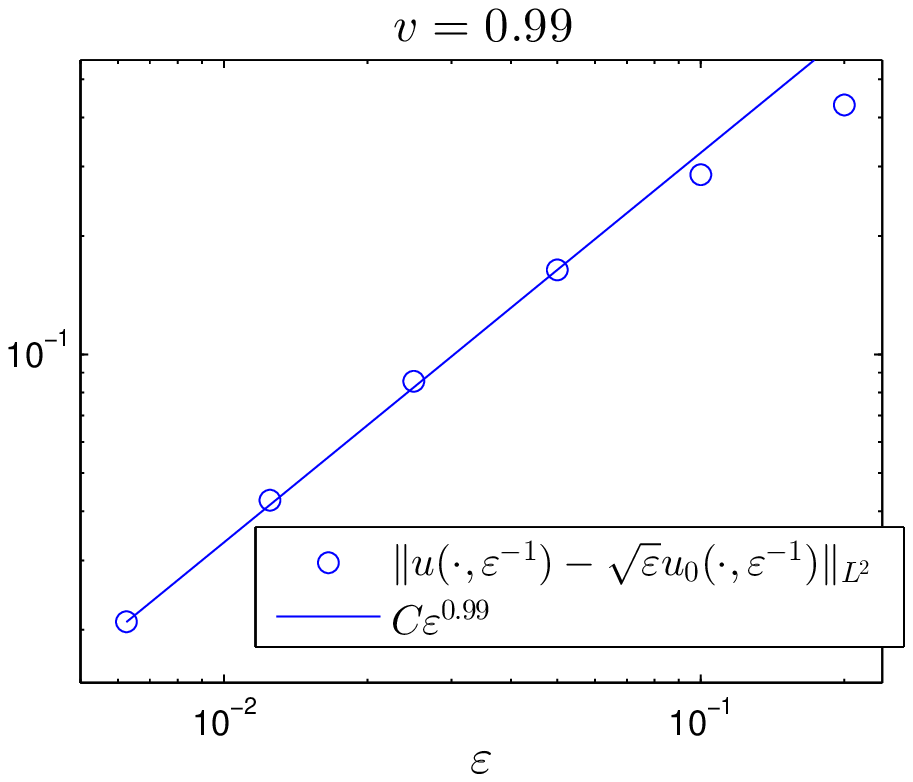,scale=0.60}
\caption{\label{F:eps_conv_cos} \small Convergence of the $L^2$-error  at $t=\eps^{-1}$, i.e. of $\|u(\cdot,\eps^{-1})-\sqrt{\eps} u_0(\cdot, \eps^{-1})\|_{L^2}$, for the choice of $\Vfb(x;1,1/2)$ given in \eqref{E:jac_ell} and $W(x)=\cos\left(\tfrac{4\pi x}{d}\right)$, $k_0=0, \omega_0\approx 3.428$. Three gap soliton velocities are tested: $v=0, 0.5,$ and $0.99$. The slopes of the full lines are obtained by interpolations of all the computed error values except for $v=0.99$, where only the four smallest values are interpolated.}
\end{center}
%sol_data_PNLS_W_cos_vel_p99_eps_p1_num_envel.mat etc.
\end{figure}

As a second example we choose
\[W(x) = \cos(4\pi x/d)\]
and study, once again, gap solitons in the gap bifurcating from the intersection of band functions at $(k_0,\omega_0) \approx  (0,3.428)$. We obtain 
\begin{align*}
&c_g \approx 3.3666546, \ \kappa \approx  0.4934402, \ \kappa_s \approx 0.0048743,\\ 
&\alpha \approx  0.1357055, \ \beta \approx -6.4865469*10^{-4}, \ \gamma \approx  -7.2277824*10^{-6}.
\end{align*}
An analogous convergence test to that in Sec. \ref{S:W_odd} produces convergence rates between $O(\eps^{0.9})$ and $O(\eps^1)$, see Fig. \ref{F:eps_conv_cos}. 

\section{Conclusions}
Using a formal asymptotic analysis and numerical tests of convergence of the approximation error with respect to the asymptotic parameter, it is shown that periodically perturbed finite band potentials of finite contrast in one dimension support a family of 
approximate moving gap solitons. The frequency parameter of the gap solitons lies in a narrow gap generated by the periodic perturbation (of amplitude $0< \eps \ll 1$). The gaps open from transversal crossings of band functions and the Bloch waves at a bifurcation point thus carry a nonzero group velocity. The problem of selection of the Bloch waves out of the two dimensional eigenspace is solved by an optimization algorithm.

While existing results on gap solitons in finite contrast structures produce only gap solitons with infinitesimal velocity, the new gap solitons propagate at an $O(1)$ velocity across the periodic structure. The convergence tests confirm the $\eps^1$-convergence rate of the $L^2$-error on time scales of $O(\eps^{-1})$ as expected from the formal asymptotics. Rigorous estimates of the approximation error will be the subject of a future paper.

\bibliographystyle{abbrv}
\bibliography{lit_fin_gap}

\end{document}